\documentclass{iopart}

\usepackage{epsf}

\begin{document}

\title{Critical temperature for entanglement transition in Heisenberg Models}

\author{Hongchen Fu$^\dagger$, Allan I Solomon$^\dagger$ and
Xiaoguang Wang$^\ddagger$}

\address{ $^\dagger$
Quantum Processes Group, The Open University, Milton Keynes, MK7
6AA, U.K.}

\address{$^\ddagger$
Department of Physics, Macquarie University, Sydney, New South
Wales 2109, Australia }


\begin{abstract}
We study thermal entanglement in some low-dimensional Heisenberg
models. It is found that in each model there is a critical
temperature above which  thermal entanglement is absent.
\end{abstract}


\section{Introduction}

Entanglement\cite{schr} plays an important role in quantum
computation and quantum information processing. With appropriate
coding, a system of interacting spins, such as described by a
Heisenberg hamiltonian,  can be used to model a solid-state
quantum computer.  It is therefore of some significance to study
thermal entanglement in Heisenberg models. We find  that for each
model  there is a corresponding  critical temperature for
transition to the entanglement regime, and the entanglement only
occurs below this critical temperature.

\section{Measures of entanglement}

A pure state described by the wave function $|\Psi\rangle$ is
{\it non-entangled} if it can be factorized as $|\Psi\rangle =
|\Psi_1\rangle \otimes |\Psi_2\rangle$. Otherwise, it is
entangled. A typical example of an  entangled state is the Bell
state for a  bipartite system of two qubits:
\begin{equation}
   \frac{1}{\sqrt{2}} (|01\rangle-|10\rangle )
\end{equation}
For such a bipartite system the most popular entanglement measure
is the {\em entanglement of formation}. For a pure state the
entanglement of formation is defined as the reduced entropy of
either subsystem\cite{wern}.

For the two-qubit system one can use {\em
concurrence}\cite{wooters} as a measure of the entanglement. Let
$\rho _{12}$ be the density matrix of the pair which may
represent either a pure or a mixed state. The concurrence
corresponding to the density matrix is defined as
\begin{equation}
{\cal C}_{12}=\max \left\{ \lambda _1-\lambda _2-\lambda
_3-\lambda _4,0\right\} ,  \label{eq:c1}
\end{equation}
where the quantities $\lambda _i$ are the square roots of the
eigenvalues of the operator
\begin{equation}
\varrho _{12}=\rho _{12}(\sigma _1^y\otimes \sigma _2^y)\rho
_{12}^{*}(\sigma _1^y\otimes \sigma_2^y)  \label{eq:c2}
\end{equation}
in descending order. The eigenvalues of $\varrho _{12}$ are real
and non-negative even though $\varrho _{12}$ is not necessarily
Hermitian. The entanglement of formation is a monotonic function
of the concurrence, whose values  range from zero, for an
non-entangled state, to one, for a maximally entangled state.

\section{Heisenberg models}

The general $N$-qubit Heisenberg XYZ model in a magnetic field
$B$ is described by the Hamiltonian
\begin{equation}
H=\frac{1}{2}\sum_{n=1}^N \left( \sigma_n^x \sigma_{n+1}^x +
\sigma_n^y \sigma_{n+1}^y + \sigma_n^z \sigma_{n+1}^z \right) +
\sum_{n=1}^N B_n \sigma_n^z
\end{equation}
where we assume cyclic boundary conditions $N+1\equiv1$. The
Gibbs state of a system in thermodynamic equilibrium is
represented by the density operator
\begin{equation}
\rho(T) = \exp(-H/kT)/Z,
\end{equation}
where $Z = \mbox{tr}[\exp(-H/kT)] $ is the partition function,
$k$ is Boltzmann's constant which we henceforth take equal to 1,
and $T$ is the temperature.

As $\rho(T)$ represents a thermal state, the entanglement in the
state is called {\em thermal entanglement}. At $T=0$, $\rho(0)$
represents the ground state which is pure for the non-degenerate
case and mixed for the degenerate case. The ground state may be
entangled. At $T=\infty$, $\rho(\infty)$ is a completely random
mixture and cannot be entangled.

\section{Thermal entanglement in the 2-site Heisenberg model}

The density matrix can be  obtained\cite{wang} as
\begin{equation}
\rho(T)=A\left(
\begin{array}{cccc}
e^{-B/T} &&& \\
& \cosh(J/T) & -\sinh(J/T) & \\
& -\sinh(J/T) & \cosh(J/T) & \\
&&& e^{B/T}
\end{array}
\right) \label{rho12}
\end{equation}
where $A=(2\cosh(J/T)+2\cosh(B/T))^{-1}$, and the concurrence
\begin{equation}
    C=\max \left\{\frac{\sinh(J/T) -1}{\cosh(J/T)+\cosh(B/T)}, 0 \right\}.
\end{equation}
As the denominator is always positive, the entanglement condition
is
\begin{equation}
    \sinh(J/T)-1>0 \qquad \mbox{or} \qquad T<1.134J.
\end{equation}
from which we conclude that
\begin{itemize}
\item There is a critical temperature $T_c \sim 1.134 J$. The thermal state
    is entangled when $T<T_c$.
\item The critical temperature is independent of the magnetic field $B$.
\item Entanglement occurs only for the antiferromagnetic case ($J>0$).
\end{itemize}

\section{Thermal entanglement in the 3-site Heisenberg model}

We now consider  pairwise entanglement in the 3-site Heisenberg
model with uniform  magnetic field, and also in the presence of a
magnetic impurity\cite{fws}. The reduced density matrix of two
sites can be written as
\begin{equation} \label{aaa}
    \rho_{12} = \frac{2}{3Z}\left(
    \begin{array}{cccc}
        u & & & \\
           & w & y & \\
           & y & w & \\
           &&& v
    \end{array}
    \right)
\end{equation}
The concurrence may be readily obtained as
\begin{equation}
    C=\frac{4}{3Z} \max \left\{ |y|-\sqrt{uv}, 0\right\},
\end{equation}

In the case of a uniform magnetic field, the entanglement between
any two sites is the same due to cyclic symmetry. We therefore
need only consider the entanglement between sites 1 and 2. Then
\begin{eqnarray}
    u(B)&=& v(-B)=\frac{3}{2} e^{3\beta B} + \frac{1}{2} e^{\beta B}(2z+z^{-2}) \nonumber \\
    w &=& \cosh(\beta B) (2z+z^{-2}) \nonumber \\
    y &=& \cosh(\beta B) (z^{-2}-z) \nonumber \\
    Z &=& 2\cosh(3\beta B) + 2\cosh(\beta B)(2z+z^{-2}).
\end{eqnarray}
where ($z=\exp(\beta J)$).

If $B=0$, one can easily find that the sites 1 and 2 are
entangled if and only if
\begin{equation}
2|z^{-2}-z|-3-2z-z^{-2} >0
\end{equation}
from which we conclude that
\begin{itemize}
\item There is no entanglement when $J>0$;
\item Entanglement occurs  when $J<0$ and $T<T_c$, where the critical temperature
is given by $-1.27J = 1.27|J|$.
\item The maximal concurrence is 1/3,  which occurs for $T\to 0$.
\end{itemize}

Fig.1  plots the concurrence against $\tau$ for different $B$.
From these graphs we see that there exists a critical temperature
above which the entanglement vanishes. It is also noteworthy that
the critical temperature increases as the magnetic field $B$
increases.
\begin{figure}
\begin{center}
\epsfxsize=6cm \epsffile{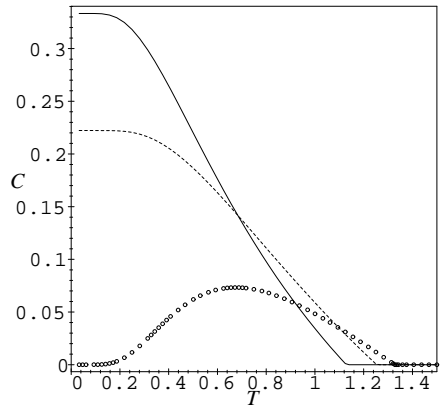}
\end{center}
\caption{ Concurrence as a function of $T$ for different magnetic
fields $B=1$(solid line), 3/2(dashed line), and 2(circle point
line). }
\end{figure}

We now consider the case of a single impurity field on the third
site; thus $B_1=B_2=0$ and $B_3 =BJ > 0$. In this case the cyclic
symmetry is violated and we have to consider the entanglement
between sites 1 and 2, and between sites 1 and 3, separately.

\begin{figure}
\begin{center}
\epsfxsize=6cm \epsffile{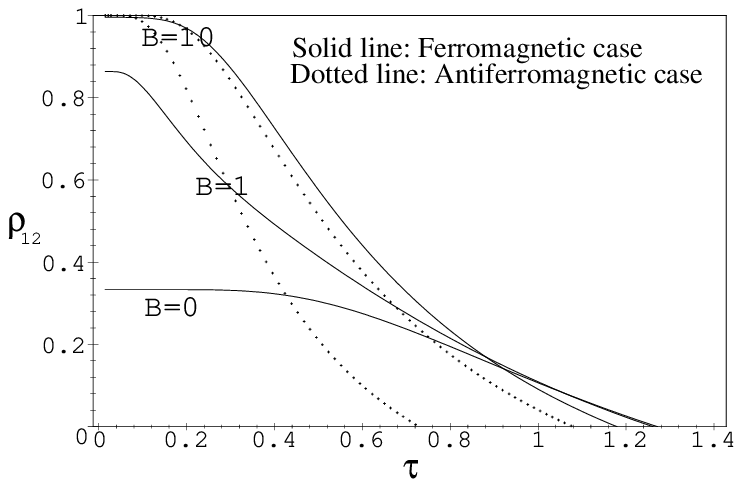} \epsfxsize=6cm
\epsffile{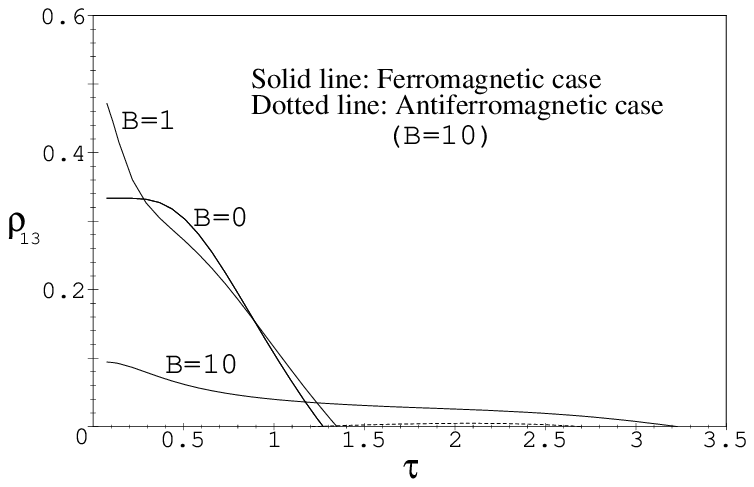}
\end{center}
\caption{Concurrence $C_{12}$ $C_{13}$against $\tau$ for
different $B$. For antiferromagnetic case (dotted line), $B=10$.}
\end{figure}

Fig.2 plots the concurrence $C_{12}$ and $C_{13}$ against scaled
temperature $\tau=kt/|J|$ for different magnetic fields $B$. From
Fig.2(a) we see that when the magnetic field is located at the
third site both the antiferromagnetic and ferromagnetic cases are
entangled in the range $0<\tau\leq \tau_c$, where the critical
temperature $\tau_c$ depends on $B$. Fig.2(a) also suggests that
the concurrence $C_{12}$ tends to 1, namely that the  $(1,2)$
entanglement becomes maximal, when $\tau\to 0$ for large enough
$B$, in both the antiferromagnetic and ferromagnetic cases.

In contrast to the $(1,2)$ case, the entanglement between sites 1
and 3 increases to a maximum with increasing $B$ and then
decreases. The lower the $\tau$, the smaller the $B$ at which the
concurrence reaches its maximum value. For smaller $B$,
entanglement occurs only in the ferromagnetic case ($J<0$), while
for  large enough $B$ (e.g. $B=10$ in our units), weak
entanglement occurs in both the antiferromagnetic and
ferromagnetic cases.


\begin{thebibliography}{99}
\bibitem{schr}Schr\"{o}dinger E 1935 Naturwissenschaften 23 807
\bibitem{wern} Werner R F 1989 Phys.Rev.A 40 4277
\bibitem{wooters} Hill S and Wootters W K 1997 Phys.Rev. Lett. 78 5022; Wootters W K 1998
Phys.Rev. Lett. 80 2245; Coffman V, Kundu J and Wootters W K 2000
Phys.Rev. A 61 052306.
\bibitem{wang}Wang X 2001 Phys.Rev. A 64 012313 Wang X 2001 Phys. Lett. A
281101
\bibitem{fws} Wang X, Fu H and Solomon A I 2001 J. Phys. A 34 11307.

\end{thebibliography}
\end{document}